\documentclass{article}
\usepackage[pdftex]{graphicx}

\usepackage[preprint]{neurips_2025}

\usepackage[utf8]{inputenc} 
\usepackage[T1]{fontenc}    
\usepackage{url}            
\usepackage{booktabs}       
\usepackage{amsfonts}       
\usepackage{nicefrac}       
\usepackage{xcolor}         

\usepackage{makecell}
\usepackage{amsmath}

\title{Improving GANs by leveraging the quantum noise from real hardware}

\author{
  Hongni Jin,  Kenneth M. Merz, Jr. \\
  Center for Computational Life Sciences, Lerner Research Institute \\
  Cleveland Clinic, Cleveland, OH 44106\\
  \texttt{\{jinh2,merzk\}@ccf.org} \\
}

\begin{document}

\maketitle

\begin{abstract}
We propose a novel approach to generative adversarial networks (GANs) in which the standard i.i.d. Gaussian latent prior is replaced or hybridized with a quantum-correlated prior derived from measurements of a 16-qubit entangling circuit. Each latent sample is generated by grouping repeated shots per qubit into a binary fraction, applying the inverse Gaussian CDF to obtain a 16-dimensional Gaussian vector whose joint copula reflects genuine quantum entanglement, and then projecting into the high-dimensional space via a fixed random matrix. By pre-sampling tens of millions of bitstrings, either from a noiseless simulator or from IBM hardware, we build large pools of independent but internally quantum-correlated latents. We integrate this prior into three representative architectures (WGAN, SNGAN, BigGAN) on CIFAR-10, making no changes to the neural network structure or training hyperparameters. The hybrid latent representations incorporating hardware-derived noise consistently lower the FID relative to both the classical baseline and the simulator variant, especially when the quantum component constitutes a substantial fraction of the prior. In addition, we execute on the QPU in parallel to not only save computing time but also further decrease the FID up to 17\% in BigGAN. These results indicate that intrinsic quantum randomness and device-specific imperfections can provide a structured inductive bias that enhances GAN performance. Our work demonstrates a practical pipeline for leveraging noisy quantum hardware to enrich deep-generative modeling, opening a new interface between quantum information and machine learning. All code and data are available at \url{https://github.com/Neon8988/GAN_QN.git}.

\end{abstract}

\section{Introduction}
Generative adversarial networks(GANs)\citep{goodfellow2014generativeadversarialnetworks} have revolutionized the field of generative modeling by framing the learning problem as a two-player game between a generator \( G \) and a discriminator \( D \). The generator network seeks to produce realistic data samples from random noise, effectively learning a mapping from a latent space to the data distribution of interest. In contrast, the discriminator network is tasked with distinguishing between genuine data samples and those synthesized by the generator. Through iterative adversarial training, the generator progressively improves its ability to produce realistic outputs, while the discriminator simultaneously enhances its discriminative capabilities\citep{gui2020reviewgenerativeadversarialnetworks,Chakraborty_2024,salehi2020generativeadversarialnetworksgans,Creswell_2018}. This adversarial training paradigm has enabled GANs to generate remarkably realistic images, videos, and audio, as well as achieve impressive performance in tasks such as image super-resolution, style transfer, and data augmentation\citep{biswas2023generativeadversarialnetworksdata,donahue2019adversarialaudiosynthesis,kreuk2023audiogentextuallyguidedaudio,ledig2017photorealisticsingleimagesuperresolution,ma2024videovideogenerativeadversarial,Zhu_2017_ICCV}.

However, GANs are notoriously challenging to train, often experiencing instability issues such as mode collapse, vanishing gradients, and convergence difficulties\citep{gulrajani2017improvedtrainingwassersteingans}. Over the past decade, several representative GAN variants have emerged, addressing these challenges through unique modifications. For example, Wasserstein GAN(WGAN)\citep{gulrajani2017improvedtrainingwassersteingans} introduces a new training objective based on the Wasserstein distance rather than the original Jensen–Shannon divergence. By enforcing Lipschitz continuity through weight clipping or gradient penalty, WGAN provides stable and meaningful gradients, mitigating issues such as mode collapse and vanishing gradients, thus significantly improving the robustness and convergence of GAN training. In addition, Spectral Normalization GAN(SNGAN)\citep{miyato2018spectralnormalizationgenerativeadversarial} introduces spectral normalization to the discriminator's weight matrices, limiting their spectral norms and thus enforcing Lipschitz continuity without heavy computational overhead. This stabilizes GAN training, substantially reduces sensitivity to hyperparameters, and effectively addresses instability problems commonly encountered in standard GAN training. Meanwhile, BigGAN\citep{brock2019largescalegantraining} utilizes large-scale training strategies combined with a highly expressive generator and discriminator architecture. Key innovations include orthogonal regularization, spectral normalization, and extensive batch sizes, which collectively enhance the training stability and produce significantly higher-quality images, capable of effectively generating diverse and realistic images at unprecedented resolutions. The StyleGAN family\citep{karras2019stylebasedgeneratorarchitecturegenerative,Karras_2020_CVPR,10.1145/3528233.3530738} adopts a novel architecture that implements style-based generation through adaptive instance normalization (AdaIN). This technique allows explicit control over the style of generated images by manipulating latent vectors at multiple layers of the network. This innovation significantly enhances image realism and resolution and enables fine-grained control over image attributes, revolutionizing the potential applications of GANs in high-resolution image synthesis. The continued evolution of GAN architectures and training techniques holds great promise for further advancements in generative modeling and its applications across various domains.

Meanwhile, quantum GANs(QGANs) have emerged at the intersection of quantum computing and generative modeling. QGANs takes advantage of the fundamental principles of quantum computing, including superposition, entanglement, and interference, to enhance the representational capacity and computational efficiency of traditional GAN architectures.\citep{Lloyd_2018} Typically, QGANs incorporate parameterized quantum circuits into their generative and discriminative models. The quantum generator is designed to prepare quantum states representing probability distributions, while the discriminator may either remain classical or similarly adopt a quantum circuit to evaluate the authenticity of generated quantum states against target quantum data distributions.\citep{Dallaire_Demers_2018} QGANs have shown promise in various applications, such as quantum state preparation, quantum simulation, and quantum data compression. For example, Hu et al.\citep{doi:10.1126/sciadv.aav2761} reported the first experimental proof-of-principle demonstration on a superconducting quantum processor, where the quantum generator learned to mimic the output statistics of a quantum channel simulator with an average fidelity of 98.8\%, effectively fooling the discriminator after several rounds of quantum-gradient updates. In addition, Patch-QGAN 
leverages inherent quantum parallelism to efficiently explore high-dimensional probability spaces, potentially enabling exponential advantages in sampling complexity and convergence speed over classical counterparts~\citep{Huang_2021}. Recently, the Patch-QGAN framework has been incorporated into photonic quantum computing, training the generator end-to-end to produce classical image samples.\citep{Sedrakyan_2024} Importantly, QuGAN runs both the generator and discriminator purely on quantum state fidelity via the SWAP test on qubit registers to achieve stable convergence with significantly reduced parameter counts and high-fidelity state reproduction.\citep{9605352} 

Despite these promising demonstrations, several key limitations persist in QGANs research. First, due to hardware and algorithmic constraints, most QGANs implementations are designed with only a few qubits, confining them to very low-dimensional feature spaces. In the current Noisy Intermediate-Scale Quantum(NISQ) era, every additional qubit exacerbates gate errors, crosstalk and decoherence, so deeper or larger circuits rapidly exceed device coherence times and causes training to fail\citep{rudolph2024trainability}. Moreover, the majority of QGANs are carried out on classical simulators, whose runtime and memory requirements grow exponentially as qubit count increases, making it infeasible to run a large quantum circuit classically. These technical limitations collectively hinder QGANs ability to explore richer, high-dimensional feature spaces. Second, given the limited expressivity of quantum circuits with only a few qubits, evaluations of QGANs have largely been restricted to toy benchmarks. For example, compressed MNIST dataset\citep{726791} with only 8x8 or 28x28 grayscale images are frequently used in QGANs experiments \citep{chang2024latentstylebasedquantumgan,shu2024variationalquantumcircuitsenhanced,9605352}, yet these simplistic benchmarks fail to capture the complexity of natural images, audio signals, or other rich data modalities. Third, in efforts to demonstrate quantum advantage, many studies\citep{chang2024latentstylebasedquantumgan, 9605352,Sedrakyan_2024} compare QGANs against classical GANs whose trainable parameters have been deliberately curtailed to only hundreds or thousands of neurons to match the depth or width of quantum circuits. Such unfairly constrained baselines significantly handicap classical models, yielding performance gaps that are inconclusive at best. Addressing these issues by scaling to richer feature sets, benchmarking on complex, real-world datasets, establishing fair classical baselines, and developing more expressive yet hardware-efficient circuit designs will be crucial to rigorously assess the true potential of QGANs.

In this work, we propose a new hybrid framework QGANs by enriching the latent prior of a classical GAN with genuine quantum-correlated noise. By sampling from a parameterized 16-qubit circuit, we obtain a white-noise prior endowed with entanglement-driven correlations. This quantum-involved noise can act as a structured inductive bias, while the classical neural generator retains the full capacity to learn the target data manifold. The result is a synergy of quantum randomness and deep-learning scalability, yielding richer latent representations and improved sample fidelity over conventional Gaussian priors. Importantly, our results show that current noisy hardware devices indeed achieve better performance than a noiseless simulator in generative modeling, indicating that the device-specific imperfections are not merely a source of error but in fact a valuable resource: the intrinsic decoherence and readout noise of real devices injects additional stochastic structure that enriches the latent prior, improving generative diversity and fidelity. In other words, hardware‐specific imperfections act as a built-in data augmentation mechanism, seeding the GAN with richer non-Gaussian correlations that a perfect simulator cannot replicate. This finding flips the notion that conventional noisy quantum hardware is a liability into an advantage for next-generation generative modeling. We further optimize our pipeline by executing multiple parameterized circuits in parallel on the quantum device. This parallel execution not only accelerates quantum computing but also increases the effective variety of quantum correlations in the latent prior, achieving an extra boost in generative performance.

\section{Quantum correlated latent representation}
In GANs the generator \(G\) and the discriminator \(D\) are trained in opposition. The generator learns to map samples from a simple latent distribution \(p_{\mathbf{z}}\) into the data space, producing synthetic examples that resemble the true data distribution \(p_{\mathrm{data}}\). In the canonical formulation, the latent prior is chosen to be a multivariate standard Gaussian,  
\[
\mathbf{z}_{\mathrm{class}} \sim \mathcal{N}(\mathbf{0}, \mathbf{I})\,,
\]  
because its form facilitates gradient‐based optimization and encourages the generator to explore all directions in latent space. During training, \(G\) aims to transform these Gaussian noise vectors into realistic data samples, while \(D\) seeks to distinguish real examples from \(G\)’s outputs. The adversarial loss  
\[
\min_{G}\,\max_{D}\; \mathbb{E}_{\mathbf{x}\sim p_{\mathrm{data}}}\bigl[\log D(\mathbf{x})\bigr]
\;+\;
\mathbb{E}_{\mathbf{z}\sim \mathcal{N}(\mathbf{0},\mathbf{I})}\bigl[\log\bigl(1 - D(G(\mathbf{z}))\bigr)\bigr]
\]  
drives both networks: \(G\) progressively “fools” \(D\), and \(D\) sharpens its discrimination, resulting in a generator capable of producing high-fidelity samples. The Gaussian prior’s isotropy and simplicity also make it straightforward to sample, interpolate, and analyze in downstream tasks. However, this simple Gaussian prior is entirely featureless because all coordinates are independent and identically distributed, leaving the network to discover any useful internal correlation from scratch. Such unstructured noise can lead to mode collapse and slow convergence, since there is no built-in bias guiding the generator toward coherent patterns. 

Unlike classical computing, quantum computing manipulates information using qubits, which can exist in coherent superpositions of 0 and 1. Importantly, qubits can become entangled, forming joint states that exhibit correlations stronger and more intricate than any classical random process. A measurement on one qubit instantaneously conditions the outcome statistics of its partner in a way that cannot be reproduced by independent or merely classically correlated variables. These genuinely quantum correlations encode non-Gaussian, nonseparable dependencies across the entire register. By sampling from an entangling circuit rather than drawing an independent Gaussian, we can inject this unique, higher-order structure directly into the GANs' latent prior, providing the generator with an inductive bias that classical noise sources simply cannot match. The quantum-correlated prior can encode non-trivial dependencies, such as the co-occurrence of textures, edges, and color channels in real images, so the generator can combine meaningful features more efficiently, cover a broader set of modes with fewer samples, and converge more rapidly than with an unstructured Gaussian prior.

To construct a quantum-correlated latent vector  \( \mathbf{z} \in \mathbb{R}^{z_{\text{dim}}} \) for GANs training, we design a 16-qubit quantum circuit which includes ECR, SX, X, Rx, Ry, Rz gates. The angles of rotational gates (Rx, Ry, Rz) are randomly initialized once and then fixed. This parameterized circuit is measured with tens of millions of shots to construct a comprehensive quantum noise pool. Specifically, the Sampler primitive in Qiskit\citep{javadiabhari2024quantumcomputingqiskit} is used to generate the bitstrings as the output of the quantum circuit. These $\sim$ 20 million 16-bit strings are stacked into a 16-dimensional latent pool, which includes 0 and 1.  To transform these binary outcomes into a mini-batch continuous prior suitable for GANs, we first group the bitstrings into blocks of \( S \) repeated measurements per qubit. We take $B \times S$ bitstrings from the pool and reshape them into a tensor 
$\mathbf{B} \in \{0,1\}^{B \times S \times 16}$. Thus each sample $b$ in the batch has its own block $\mathbf{B}_{b,:,:}$ of $S$ repeated measurements on every qubit, guaranteeing independence across samples.

For each sample $b$ and qubit $q$, we interpret the $S$ bits $\{b_{s,q}\}$ as the binary fraction:

\begin{equation}
u_{b,q}
= \sum_{s=1}^{S} b_{s,q}\,2^{-(s+1)}
\;\in\;[0,1].
\label{eq:ubq}
\end{equation}

We then apply the inverse normal CDF, $x_{b,q} = \Phi^{-1}(u_{b,q})$ to transform the discrete measurement data into a continuous, zero-mean, unit-variance latent 
$\mathbf{X} \in \mathbb{R}^{B \times 16}$
while still preserving the entanglement-driven correlation structure. Next, a fixed, row-orthonormal matrix $\mathbf{P} \in \mathbb{R}^{z_{\text{dim}} \times 16}$ maps $\mathbf{X}$ into the GAN latent space:

\begin{equation}
\mathbf{z}_{\mathrm{quant}}
= \mathbf{X}\,\mathbf{P}^{T}
\;\in\;\mathbb{R}^{B \times z_{\mathrm{dim}}}\,.
\label{eq:zquant}
\end{equation}

To study the trade-off between structure and isotropy, we propose a hybrid scheme to leverage the advantages of both types of latent representations,

\begin{equation}
  \mathbf{z}_{\mathrm{hyb}}
  = \alpha\,\mathbf{z}_{\mathrm{quant}}
  + (1 - \alpha)\,\mathbf{z}_{\mathrm{class}},
  \quad \alpha \in [0,1]
  \label{eq:hybrid-prior}
\end{equation}

without renormalizing so that the off-diagonal terms from $\sum$ persist whenever $\alpha < 1$.

\begin{figure}[htbp]
    \centering
    \includegraphics[width=1\textwidth]{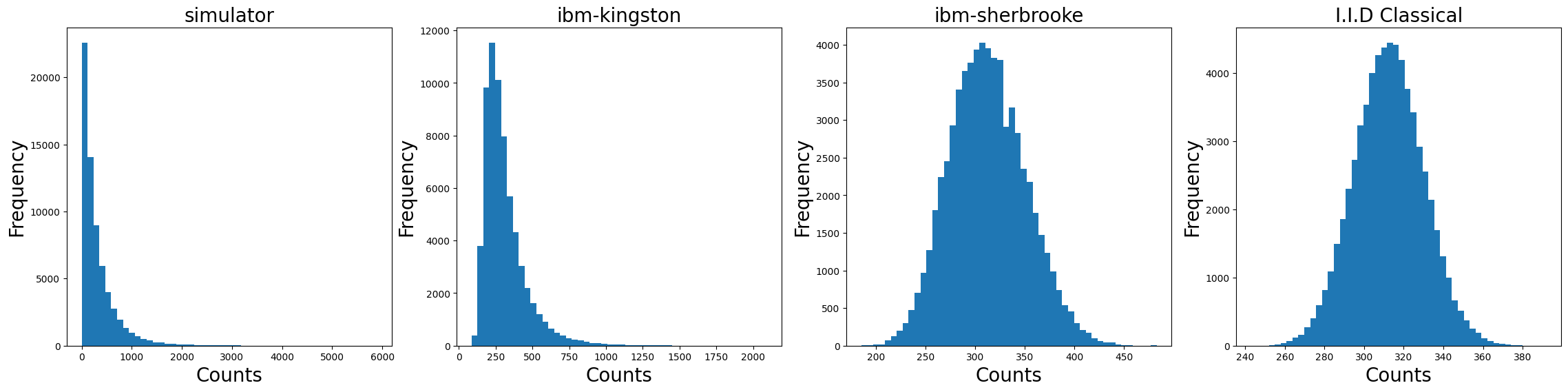}
    \caption{The bistrings frequency distribution across multiple devices. }
    \label{fig:figure1}
\end{figure}

\begin{figure}[htbp]
    \centering
    \includegraphics[width=1\textwidth]{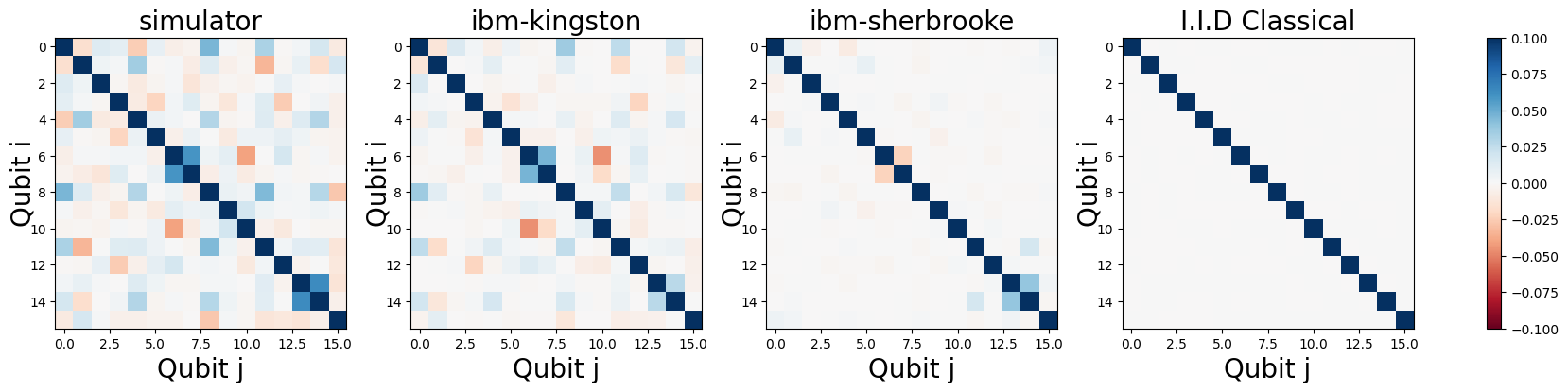}
    \caption{The qubit correlation analysis across multiple devices. }
    \label{fig:figure2}
\end{figure}

\section{Experiments}

\subsection{Bitstrings generation}
To investigate the effects of physical noise from hardware on the generated bitstrings distribution, we run the same 16-qubit parameterized circuit on multiple devices, including a noiseless simulator, ibm-sherbrooke(Eagle r3 processor) and ibm-kingston(Heron r2 processor). The Heron r2 processor generally has a lower error rate than the Eagle r3 processor, thus generating bitstrings closer to those from a noiseless simulator.  In each case, we run the circuit with 20,000 shots and repeat 1024 times to get a total of 20,480,000 bitstrings. To validate whether quantum correlations contribute to generative modeling, we also generate a bitstring pool classically by randomly sampling 0 or 1. 

Figure~\ref{fig:figure1} shows the empirical frequency distribution of our 16-qubit measurement outcomes across multiple devices. On the noiseless simulator, the circuit's deterministic biases dominate and certain bitstrings appear more than 20k times, while 248 possible patterns never occur in this large pool with $\sim$20 million bitstrings. The real device noise on ibm-kingston flattens this distribution which now features a lower peak and a longer tail of rarer bitstrings, reflecting moderate decoherence and readout errors that partially break the simulator's strong repetition. On ibm-sherbrooke where the hardware noise is higher, the histogram becomes nearly uniform, the peak frequency is low and all 65536 patterns appear with comparable counts, demonstrating the physical imperfections can vastly increase sampling diversity. Meanwhile, the uniform classical bistrings distribution confirms truly i.i.d. behavior in classical computing. 

Figure~\ref{fig:figure2} visualizes the pairwise correlation of 16-qubit bitstring outcomes across four noise sources. In the noiseless simulator, the entangling gates produce diverse strong off-diagonal entries, resulting in highly structured dependencies. On the ibm-kingston backend, physical decoherence and readout errors weaken most of these simulator-induced correlations, leaving only some medium-strength links. The ibm-sherbrooke backend further dampens the joint distribution due to the significant and unavoidable physical noise and only a few residual off-diagonal spots remain, indicating that hardware noise has largely randomized the qubit relationships while preserving minimal quantum structure. By contrast, i.i.d classical bitstrings exhibit a nearly blank off-diagonal, as expected for independent sampling. These heatmaps confirm that real-device noise erodes, but does not completely destroy entanglement-driven correlations. 

In the context of GAN priors, neither the extreme of pure randomness with uniform bitstring distribution nor rigid entanglement alone is ideal. On one hand, we need a broad, near-uniform spread of bitstrings to ensure the generator sees a rich variety of latent representation. On the other hand, we also want structured correlations, i.e., the non-trivial, entanglement-driven dependencies that encodes useful feature relationships (e.g. co-occuring edges or color patterns) as an inductive bias. A purely i.i.d. prior(classical) maximizes diversity but offers no guidance, forcing the network to learn everything from scratch. A noiseless quantum simulator delivers strong correlations but collapses into a small set of repeated patterns,which hinders GAN generalization since the model obtains mostly redundant latent representation. Ideally, a diverse bitstring distribution with certain quantum copula structures will result in diverse quantum-correlated priors which can improve GANs.

\begin{figure}[htbp]
    \centering
    \includegraphics[width=1\textwidth]{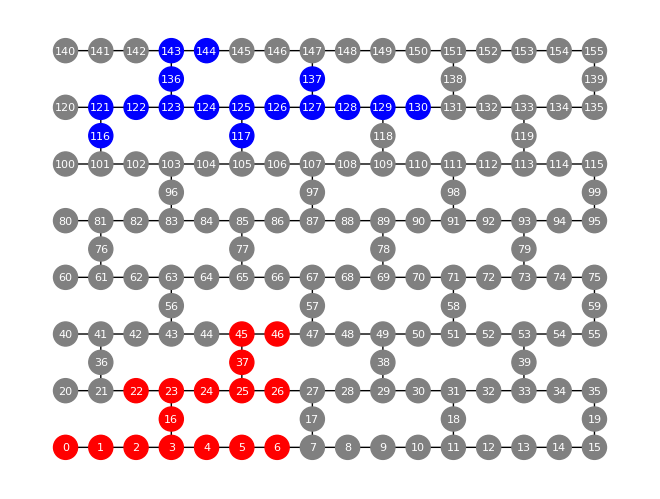}
    \caption{The physical layout of ibm-kingston. Red: layout 1(low error rate), blue:layout 2(high error rate).}
    \label{fig:figure3}
\end{figure}

\begin{table}[ht]
\caption{Two‐qubit gate depths and mean/std of the observed correlations on various devices.}
\label{tab:device_correlations}
\centering
\begin{tabular}{lccrr}
\toprule
Device            & \# Gates & Depth of 2Q Gate & \multicolumn{2}{c}{Correlation} \\
                  &          &                  & Mean   & Std    \\
\midrule
Simulator         & 564      & 65               & 0.0093 & 0.0148 \\
IBM–Kingston      & 1335     & 58               & 0.0054 & 0.0099 \\
IBM–Sherbrooke    & 1952     & 58               & 0.0016 & 0.0047 \\
QPU in Parallel   & 4891     & 259              & 0.0019 & 0.0052 \\
I.I.D Classical     & --       & --               & 0      & 0      \\
\bottomrule
\end{tabular}
\end{table}

\subsection{QPU execution in parallel}
\begin{figure}[htbp]
    \centering
    \includegraphics[width=1\textwidth]{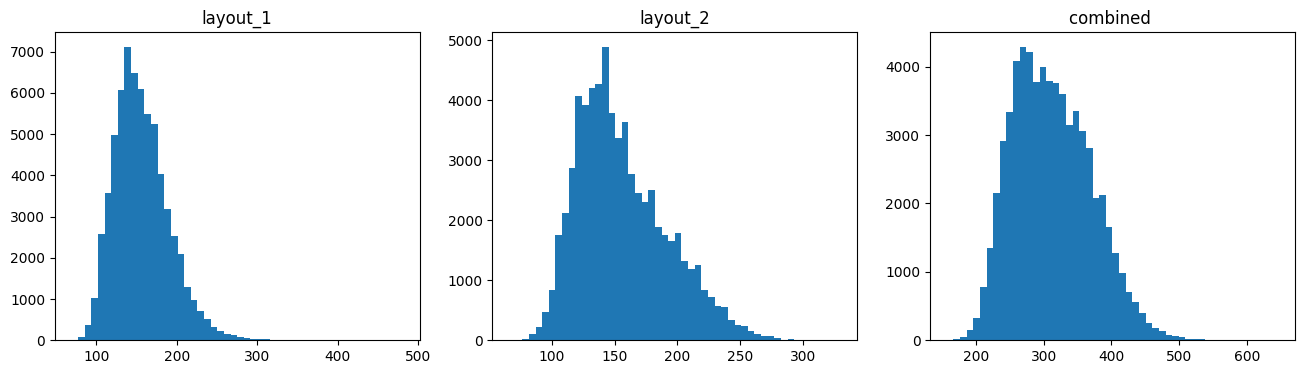}
    \caption{The bitstring frequency distribution for QPU execution in parallel on ibm-kingston. }
    \label{fig:figure4}
\end{figure}

\begin{figure}[htbp]
    \centering
    \includegraphics[width=1\textwidth]{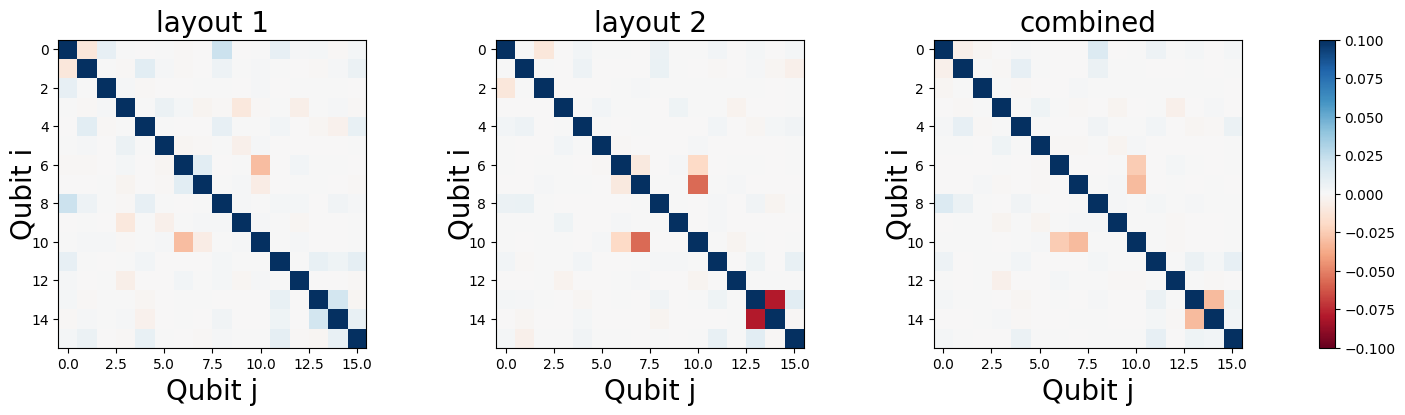}
    \caption{The qubit correlation analysis for QPU execution in parallel on ibm-kingston.}
    \label{fig:figure5}
\end{figure}

To maximize both diversity and residual entanglement in our quantum-driven GANs priors, we introduce a dual-layout QPU execution strategy. Rather than running a single 16-qubit circuit in isolation, leaving much of the hardware idle, we instantiate two independent quantum registers (and their corresponding classical registers) on the same device and transpile them with distinct physical‐layout mappings. Layout choice in Qiskit directly impacts gate error rates, a clean mapping(ibm-kingston) preserves strong qubit correlations but yields a narrow, peaked bitstring distribution, whereas a noisy mapping(ibm-sherbrooke) randomizes outcomes into a near-uniform histogram at the cost of weakened correlations(see Figure~\ref{fig:figure1}–~\ref{fig:figure2}). To reach a richly correlated latent pool with maximal entropy, we purposely assign two distinctive layouts in parallel execution, where one layout is with a high error rate to generate diverse bitstrings while another layout is assigned with a low error to maintain quantum correlations. We use mapomatic\citep{PRXQuantum.4.010327} to evaluate the error rates of all possible layout candidates on ibm-kingston and choose two appropriate layouts that maximize their error-rate disparity and whose physical qubits lie far apart to minimize crosstalk. Figure ~\ref{fig:figure3} visualizes the layouts we use for QPU execution in parallel. By combining these two half bistring pools, we expect to get a complete latent pool as Figure~\ref{fig:figure1} but with optimal tradeoff between bitstring distribution and quantum correlations. Figure~\ref{fig:figure4} shows that the merged bitstring distribution is near-uniform with a wide peak, indicating much diversity exists when sampling noise from this latent pool for GAN training. In addition, the correlation heatmap in Figure~\ref{fig:figure5} shows that stronger quantum correlations exist in this latent pool than those from ibm-sherbrooke in Figure~\ref{fig:figure2} and can also be confirmed from the quantitative analysis in Table~\ref{tab:device_correlations}.  This dual-layout approach therefore realizes the optimal trade-off: rich diversity plus certain quantum structure as latent priors.

\subsection{Image generation}
To extensively evaluate the effectiveness of quantum-involved priors in GANs, we introduce the hybrid latent prior in equation~\eqref{eq:hybrid-prior} to three representative GANs implemented in StudioGAN\citep{kang2023studiogan}: WGAN\citep{gulrajani2017improvedtrainingwassersteingans}, SNGAN\citep{miyato2018spectralnormalizationgenerativeadversarial},and BigGAN\citep{brock2019largescalegantraining}. We trained each model on the CIFAR-10 dataset which includes 60,000 32x32 images in 10 classes. For a fair comparison, all hyperparameters are kept identical to the baseline model and the only difference is how much quantum-correlated latent priors is injected to the GANs during training. The Fréchet Inception Distance (FID)\citep{heusel2017gans} is used as the benchmark metric. For all experiments, we compute the FID score with 50,000 generated images using the InceptionV3-tf backend in StudioGAN\citep{kang2023studiogan}. Each model is trained on one Nvidia A100 GPU. And for each case, we run three independent trials.

\begin{table}[ht]
\centering
\caption{FID scores (mean ± std) for different GAN models across devices. Lower is better.}
\label{tab:gan_fid_comparison}
\scalebox{0.85}{%
\begin{tabular}{lcccccc}
\toprule
Model   & Baseline      & Simulator     & IBM–Sherbrooke & IBM–Kingston  & I.I.D Classical & QPU in Parallel \\
\midrule
WGAN    & 22.81\,$\pm$\,0.40 & 21.96\,$\pm$\,0.25 & 21.50\,$\pm$\,0.17 & 21.34\,$\pm$\,0.10 & 22.12\,$\pm$\,0.43  & 21.30\,$\pm$\,0.04 \\
SNGAN   &  5.87\,$\pm$\,0.05 &  5.91\,$\pm$\,0.09 &  5.66\,$\pm$\,0.15 &  5.79\,$\pm$\,0.13 &  5.88\,$\pm$\,0.10 &  5.68\,$\pm$\,0.10 \\
BigGAN  &  4.06\,$\pm$\,0.06 &  3.51\,$\pm$\,0.06 &  3.45\,$\pm$\,0.10 &  3.82\,$\pm$\,0.19 &  3.69\,$\pm$\,0.24 &  3.37\,$\pm$\,0.16 \\
\bottomrule
\end{tabular}}
\end{table}

\begin{figure}[htbp]
    \centering
    \includegraphics[width=1\textwidth]{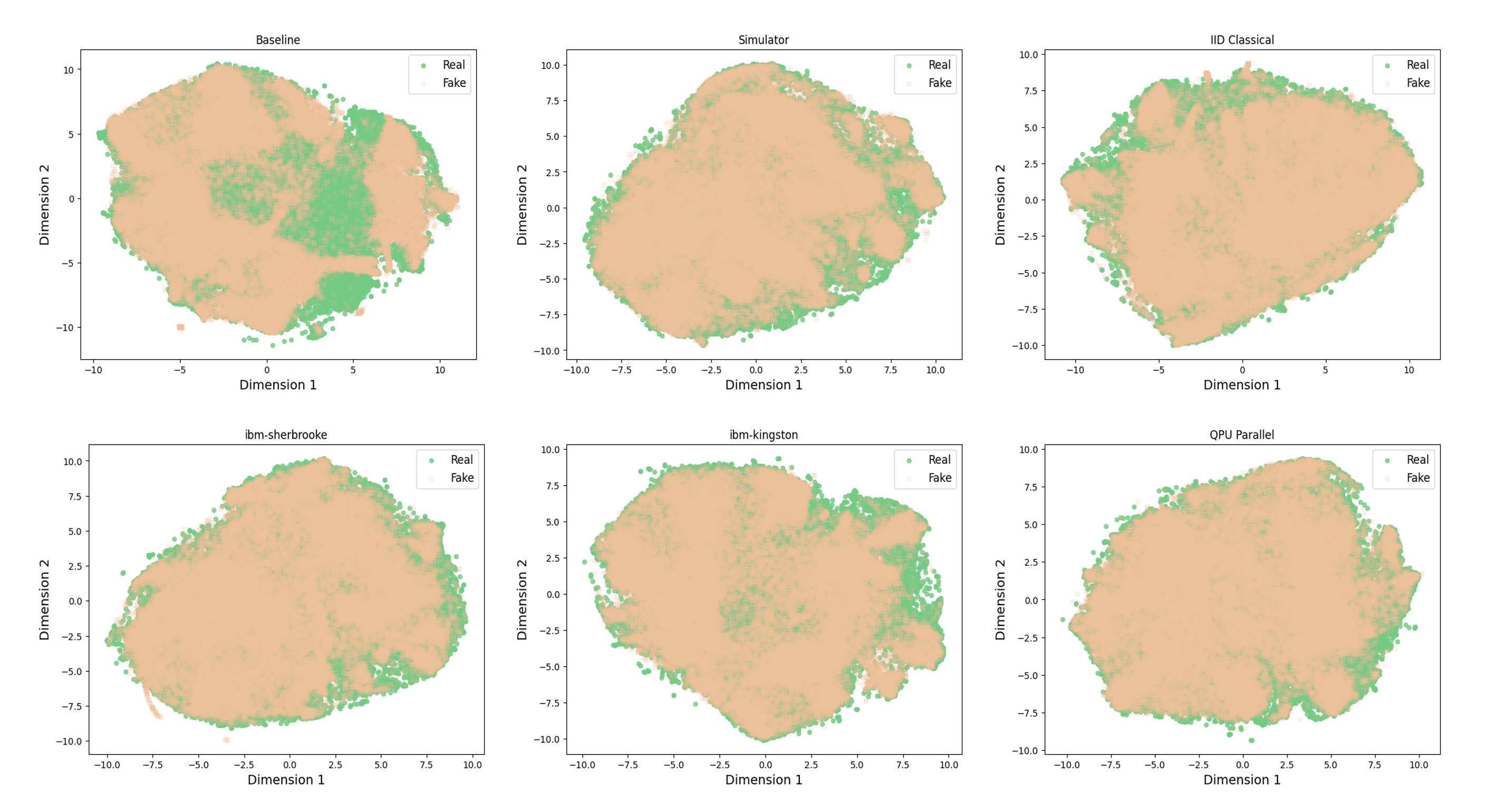}
    \caption{The t-SNE analysis across multiple devices in BigGAN.}
    \label{fig:figure6}
\end{figure}

\section{Results}
\begin{figure}[htbp]
    \centering
    \includegraphics[width=0.8\textwidth]{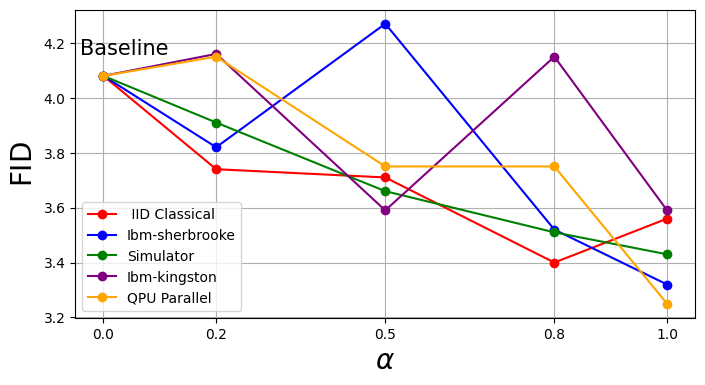}
    \caption{FID scores with different $\alpha$ across multiple devices in BigGAN.}
    \label{fig:figure7}
\end{figure}

Table~\ref{tab:gan_fid_comparison} shows that our quantum-hybrid priors yield consistent FID improvements over the classical Gaussian baseline across all three GAN architectures and that the best performance arises when we execute noisy and clean layouts in parallel on the QPU. For WGAN, simply switching to the noiseless simulator decreases FID from 22.81 to 21.96, while real hardware further reduces it to 21.50(ibm-sherbrooke) and 21.34(ibm-kingston). The QPU-parallel approach yields the lowest FID with a 7.09\% improvement compared to baseline. In SNGAN, hardware noise again outperforms the noiseless simulator, and ibm-sherbrooke yields 5.66, a 3.58\% reduction of FID, with QPU-in-parallel matching ibm-sherbrooke but with lower variance. The largest improvement is observed in BigGAN with a 17.20\% improvement from QPU-parallel. In all experiments, QPU in parallel clearly outperforms IID classical, while both have similar near-uniform bistring distributions, indicating that the extra improvement is attributed to the unique quantum correlations existing in the bistrings from quantum computing. In addition, Figure~\ref{fig:figure6} shows the t-SNE analysis by visualizing the discriminator's penultimate-layer features for both real images and generated samples. The largest gap is observed from standard Gaussian priors,followed by ibm-kingston and i.i.d classical, while ibm-sherbrooke and simulator display similar overlap, and QPU in parallel clearly shows minimal mismatch. This trend mirrors our FID results and visually confirms that the dual-layout quantum prior yields generated images whose feature distributions best match the real data manifold. Figure~\ref{fig:figure7} shows how FID changes as we adjust the hybrid mixing coefficient $\alpha$ in equation~\eqref{eq:hybrid-prior} from a pure Gaussian prior($\alpha=0$ ) to a fully quantum-correlated prior($\alpha=1$ ).  Each case displays a generally decreasing trend with $\alpha$. Importantly, in both simulator and real hardware, injecting purely quantum-correlated priors yields the lowest FID, indicating that quantum computing introduces a structured covariance that the generator can exploit.

\section{Conclusion}
In this work, we have demonstrated that seeding GANs with a quantum-correlated latent prior, derived from measurements of a 16-qubit entangling circuit, can offer a meaningful inductive bias beyond classical white noise. By mapping bitstrings into Gaussian marginals via binary-fraction encoding and inverse CDF, and then projecting into the GAN latent space, we preserve genuine quantum-induced covariances within each sample while maintaining independence across the batch. Extensive experiments on CIFAR-10 with three representative architectures (WGAN, SNGAN, BigGAN) show that hybrid or fully quantum priors yield consistent improvements in FID, especially when the quantum component is sourced from real hardware, indicating that the additional entropy and device-specific imperfections introduce a structured statistical bias that the generator can exploit. To push further, we introduced a dual-layout QPU-in-parallel execution strategy: by dispatching one circuit on a “clean” low-error mapping and a “noisy” high-error layout and then concatenating their outputs, we synthesize a latent prior that combines maximal sampling diversity with residual entanglement-driven structure, leading to the lowest FID across multiple GANs architectures. 
Our findings suggest that even with current noisy quantum processors, their intrinsic randomness can be harnessed to enhance generative modeling. Future work includes exploring larger circuits and datasets, optimizing the choice of mixing weight \( \boldsymbol{\alpha} \), and investigating theoretical links between quantum copulas and data distributions. This study opens a novel avenue at the intersection of quantum information and deep generative learning, showing that unpredictability of quantum hardware, traditionally seen as a limitation, can in fact benefit classical AI systems.

\section*{Acknowledgements}
The authors gratefully acknowledge financial support from the NIH (GM130641). We also extend our sincere gratitude
to Thaddeus Pellegrini at IBM Quantum for helpful discussion on QPU in parallel implementation.
\bibliographystyle{unsrtnat}
\bibliography{neurips_2025}

\newpage
\appendix

\renewcommand{\thefigure}{A.\arabic{figure}}
\setcounter{figure}{0}


\begin{figure}[htbp]
    \centering
    \includegraphics[width=0.6\textwidth]{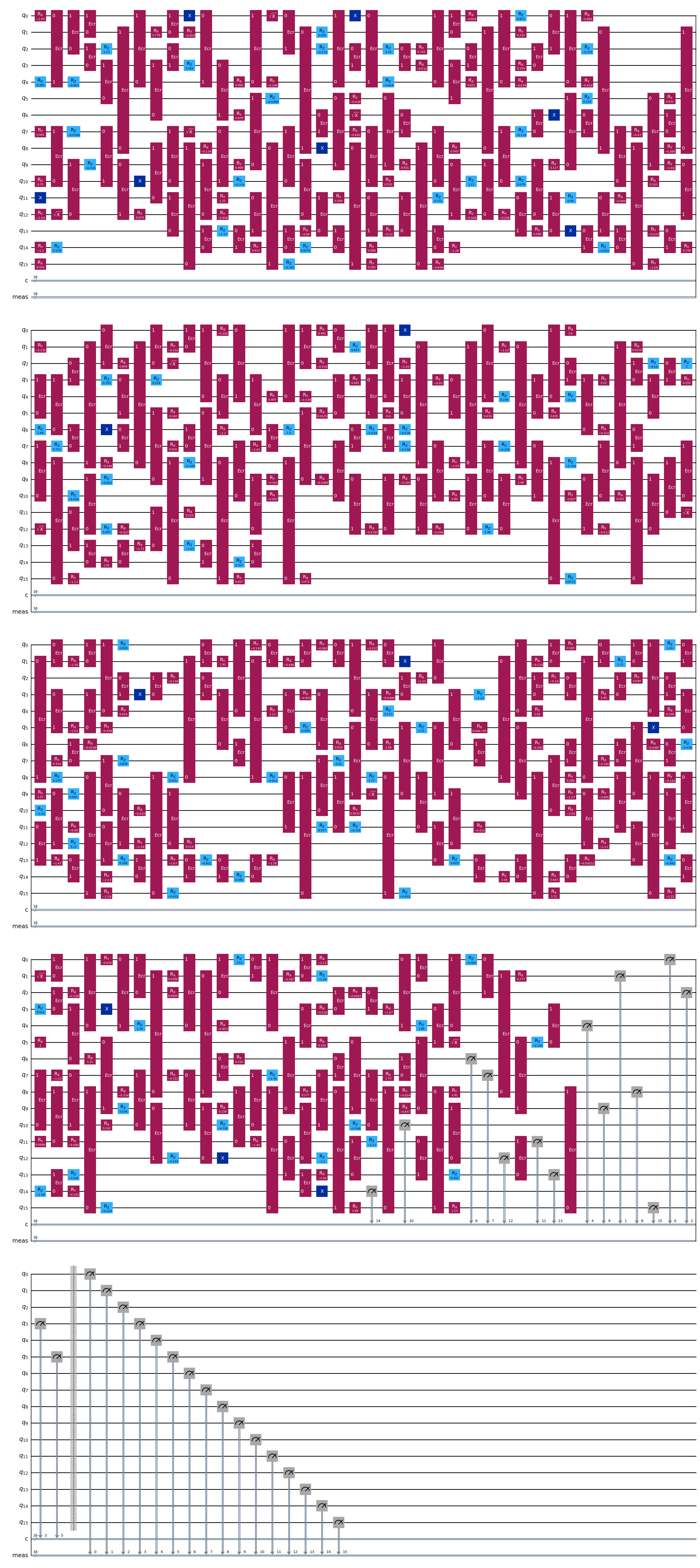}
    \caption{The 16-qubit parameterized circuit used to generate quantum-correlated bistrings.}
    \label{fig:figure a1}
\end{figure}

\begin{figure}[htbp]
    \centering
    \includegraphics[width=0.8\textwidth]{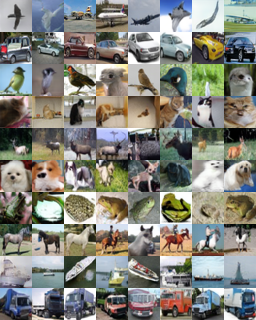}
    \caption{Randomly generated examples of images using BigGAN with quantum-involved priors by executing QPU in parallel.}
    \label{fig:figure a2}
\end{figure}

\newpage


\end{document}